%% file: few_body_bound_states.tex
\documentclass[twocolumn,english,superscriptaddress,prl]{revtex4-1}
\usepackage[T1]{fontenc}
\usepackage[utf8]{inputenc}
\PassOptionsToPackage{hyphens}{url}\usepackage{hyperref}

\usepackage{bm,dcolumn}
\usepackage{amsmath,amsfonts,amssymb,amsthm,amscd} 
\usepackage{amsxtra}

\usepackage{graphicx}
\usepackage{verbatim}
\usepackage[section]{placeins}

\usepackage[usenames, dvipsnames]{color}
\usepackage{xcolor}

\usepackage{soul} 	%

\usepackage[normalem]{ulem}

\usepackage{soul}
\usepackage{xparse}
\usepackage{physics}
\usepackage{siunitx}
\usepackage{xspace}

   \setul{0.5ex}{0.1ex}
       \definecolor{Blue}{rgb}{0,0.0,1}
    \setulcolor{Blue}

\newcommand\reduline{\bgroup\markoverwith
{\textcolor{orange}{\rule[-.5ex]{2pt}{0.4pt}}}\ULon}

\newcommand{\new}[1]{#1}

\newcommand{\cmntc}[1]{[Cmnt: \textcolor{brown}{ #1} ]}
\newcommand{\todoc}[1]{[TODO: \textcolor{gray}{#1}] }

\newcommand{\rs}{\rm \scriptscriptstyle}
\newcommand{\remove}[1]{}
\newcommand{\footnotePB}[1]{\footnote{#1}}

\graphicspath{ {./} }

\newcommand{\blst}{\begin{lstlisting}}
\newcommand{\est}{\end{lstlisting}}

\newcommand{\bit}{ \begin{itemize}{}  }
\newcommand{\eit}{\end{itemize}}

\newcommand{\be}{\begin{enumerate} \itemsep -4pt  }
\newcommand{\ee}{\end{enumerate}}

\newcommand{\bi}{\begin{itemize}   } %
\newcommand{\ei}{\end{itemize}}

\newcommand{\bma}{\begin{math}} 
\newcommand{\ema}{\end{math}}

\newcommand{\bc}{\begin{columns}} 
\newcommand{\ec}{\end{columns}}

\newcommand{\bbl}{\begin{block}} 
\newcommand{\ebl}{\end{block}}

\newcommand{\bflsh}{\begin{flashcard}}
\newcommand{\eflsh}{\end{flashcard}}

\newcommand{\bfl}[2]{\begin{flashcard}{#1} {#2} \eflsh}

\newcommand{\beq}{\begin{equation} \vspace{-0em}} 
\newcommand{\eeq}{\vspace{-0.em} \end{equation}}

\newcommand{\beqs}{\begin{equation*}}
\newcommand{\eeqs}{\end{equation*}}

\def\beqa{\begin{eqnarray}}
\def\eeqa{\end{eqnarray}}

\newcommand{\beal}{\begin{align}}
\newcommand{\eeal}{\end{align}}

\newcommand{\bvr}{\begin{verbatim}}
\newcommand{\evr}{\end{verbatim}}

\newcommand{\ra}{\ensuremath{\rightarrow}\xspace}

\newcommand{\figref}[1]{FIG.~\ref{#1}}

\newcommand{\integral}[1]{\int \! \mathrm{d} #1\,}                    %

\def\br{{\bf r}}

\def\e{\ensuremath{\mathcal{E}}\xspace}

\def\I{\ensuremath{{I}}\xspace}
\def\P{\ensuremath{{P}}\xspace}
\def\E{\ensuremath{\mathcal{E}}\xspace}

\def\nn{\nonumber}

\def\D{\ensuremath{\Delta}\xspace}

 \renewcommand{\figref}[1]{Fig.~\ref{#1}}
\renewcommand{\eqref}[1]{Eq.~(\ref{#1})\xspace}
\newcommand{\details}[1]{}

\def\pSt{\ensuremath{P}}
\def\rSt{\ensuremath{S}}

\renewcommand{\ket}[1]{\ensuremath{\left|#1\right\rangle}}  							 	%
\renewcommand{\cmntc}[1]{\footnotePB{\textbf{cmnt:} #1}}
\renewcommand{\cmntc}[1]{}
\renewcommand{\paragraph}[1]{{\it #1.---}}
\newcommand{\removeAuthor}[1]{ 
\author{Authors}
\affiliation{Joint Quantum Institute, National Institute of Standards and Technology and the University of Maryland, College Park, Maryland 20742 USA}
\affiliation{Joint Center for Quantum Information and Computer Science, National Institute of Standards and Technology and the University of Maryland, College Park, Maryland 20742 USA}
\affiliation{Other}}

\renewcommand{\removeAuthor}[1]{#1}
\def\OD{\text{OD}\xspace}

\def\D{\Delta}
\def\Foerster{F{\"o}rster\xspace}

\def\be{\begin{equation}}
\def\ee{\end{equation}}
\def\kinetic{\ensuremath{T\xspace}}
\renewcommand{\todoc}[1]{}  

\def\lineconfiguration{linear configuration\xspace}

\def\z2{z'}

\def\rEIT{Rydberg-EIT\xspace}
\def\newparagraph{} %

\newcommand\psiOpDr[1]{\hat{#1^\dagger}}
\newcommand\psiOp[1]{\hat{#1}}

\newcommand{\colthree}[3]{\ensuremath{\begin{pmatrix}#1\\#2\\#3\end{pmatrix}}}

\def\Dpol{\ensuremath{D}}

\def\Fpol{\ensuremath P}

\def\molecularlike{Lennard-Jones-like\xspace}
\usepackage{bm}%

\newcommand{\spaceCutA}[1]{\textcolor{gray}{#1}}
\newcommand{\spaceCut}[1]{\textcolor{gray}{#1}}
\renewcommand{\spaceCutA}[1]{}
\renewcommand{\spaceCut}[1]{#1}

\begin{document}
\title {Exotic photonic molecules via Lennard-Jones-like potentials}
\removeAuthor
{
\author{Przemyslaw Bienias}
\affiliation{Joint Quantum Institute, NIST/University of Maryland, College Park, Maryland 20742 USA}
\affiliation{Joint Center for Quantum Information and Computer Science, NIST/University of Maryland, College Park, Maryland 20742 USA}

\author{Michael J. Gullans}
\affiliation{Joint Quantum Institute, NIST/University of Maryland, College Park, Maryland 20742 USA}
\affiliation{Joint Center for Quantum Information and Computer Science, NIST/University of Maryland, College Park, Maryland 20742 USA}
\affiliation{Department of Physics, Princeton University, Princeton, New Jersey 08544 USA}

\author{Marcin Kalinowski}
\affiliation{Joint Quantum Institute, NIST/University of Maryland, College Park, Maryland 20742 USA}
\affiliation{Faculty of Physics, University of Warsaw, Pasteura 5, 02-093 Warsaw, Poland}

\author{Alexander N. Craddock}
\affiliation{Joint Quantum Institute, NIST/University of Maryland, College Park, Maryland 20742 USA}

\author{Dalia P. Ornelas-Huerta}
\affiliation{Joint Quantum Institute, NIST/University of Maryland, College Park, Maryland 20742 USA}

\author{S. L. Rolston}
\affiliation{Joint Quantum Institute, NIST/University of Maryland, College Park, Maryland 20742 USA}

\author{J.V. Porto}
\affiliation{Joint Quantum Institute, NIST/University of Maryland, College Park, Maryland 20742 USA}

\author{Alexey V. Gorshkov}
\affiliation{Joint Quantum Institute, NIST/University of Maryland, College Park, Maryland 20742 USA}
\affiliation{Joint Center for Quantum Information and Computer Science, NIST/University of Maryland, College Park, Maryland 20742 USA}
}

\date{\today}

\begin{abstract}
Ultracold systems offer an unprecedented level of control of interactions between atoms.  An  important challenge is to achieve a similar level of control of the interactions between photons. Towards this goal, we propose a realization of a novel Lennard-Jones-like potential between photons  coupled to the Rydberg states via electromagnetically induced transparency (EIT).  This potential is achieved by tuning Rydberg states to a F{\"o}rster resonance with other Rydberg states. We consider few-body problems in 1D and 2D geometries and show the existence of self-bound clusters (``molecules'') of photons. We demonstrate that for a few-body problem, the multi-body interactions have a significant impact on the geometry of the molecular ground state. This leads to phenomena without counterparts in conventional systems: For example,  three photons in 2D preferentially arrange themselves in a line-configuration rather than in an equilateral-triangle configuration. Our result opens a new avenue for studies of many-body phenomena with strongly interacting photons. 
\end{abstract}

\maketitle
Atomic, molecular, and optical platforms %
allow for precise control and wide-ranging tunability of system parameters using external fields.
Interactions between \textit{atoms} can be controlled via Feshbach resonances enabling studies of the BCS-BEC crossover~\cite{Bloch2008,Chin2010}  
\cmntc{, realization of strong interactions between individual photons~\cite{Chang2014}}
or via highly-excited Rydberg states 
giving rise to frustrated magnetism~\cite{VanBijnen2015}, topological order~\cite{Glaetzle2015}, %
and other exotic phases~\cite{Grass2018}.
Interactions between single \textit{photons} in vacuum or conventional  transparent materials are negligible; %
however, they can be enhanced by strongly coupling photons to specially engineered matter~\cite{Chang2014}. %
An open challenge is to achieve a similar level of tunability for strongly interacting photons as demonstrated for atoms.
Such tunability could lead to applications in photonic quantum information processing, metrology, sensing, as well as exotic photonic phases of matter~\cite{Chang2015}. 
\cmntc{This potential can give rise to shape resonances as was proposed in Ref.~\cite{Bienias2014} (corresponding to the Feschbach resonances between ultra-cold atoms).}
A promising platform %
to achieve this goal are Rydberg polaritons, for which interactions between Rydberg states are mapped onto photons via EIT~\cite{Fleischhauer2000,Lukin2001,Friedler2005}.
The effective interactions between photons are not only strong~\cite{Pritchard2010}, but also saturate to a constant value~\cite{Firstenberg2013,Bienias2014,Bienias2016a,Bienias2020} for distances shorter than the blockade radius $r_b\sim10\,\mu$m, %
which usually is much greater than the wavelength of the photons.
These properties have enabled several theoretical proposals and experimental realizations related to quantum information processing, such as gates~\cite{Tiarks2019,Tiarks2016}, transistors~\cite{Tiarks2014,Gorniaczyk2014,Gorniaczyk2016}, and non-classical states of light~\cite{Dudin2012,Peyronel2012,OrnelasHuerta2020,Bienias2020c}. %

Up to now, the field of dispersive \rEIT has predominantly concentrated on the %
 effective  interactions between polaritons proportional to $ 1/(r_b^6+r^6)$, which change monotonically as a function of separation~$r$.  
Additional work has explored: lossy~\cite{foot1} 
Coulomb bound states; and bound states via interactions mediated by 1D photonic crystals~\cite{Douglas2016}. 
%
%
Here, we propose a novel method to tune the shape %
of the divergence-free interactions~\cite{Maghrebi2015c} in 1D and 2D between photons propagating through the Rydberg medium. %
\begin{figure}[h!]%
\includegraphics[width= 0.99\columnwidth]{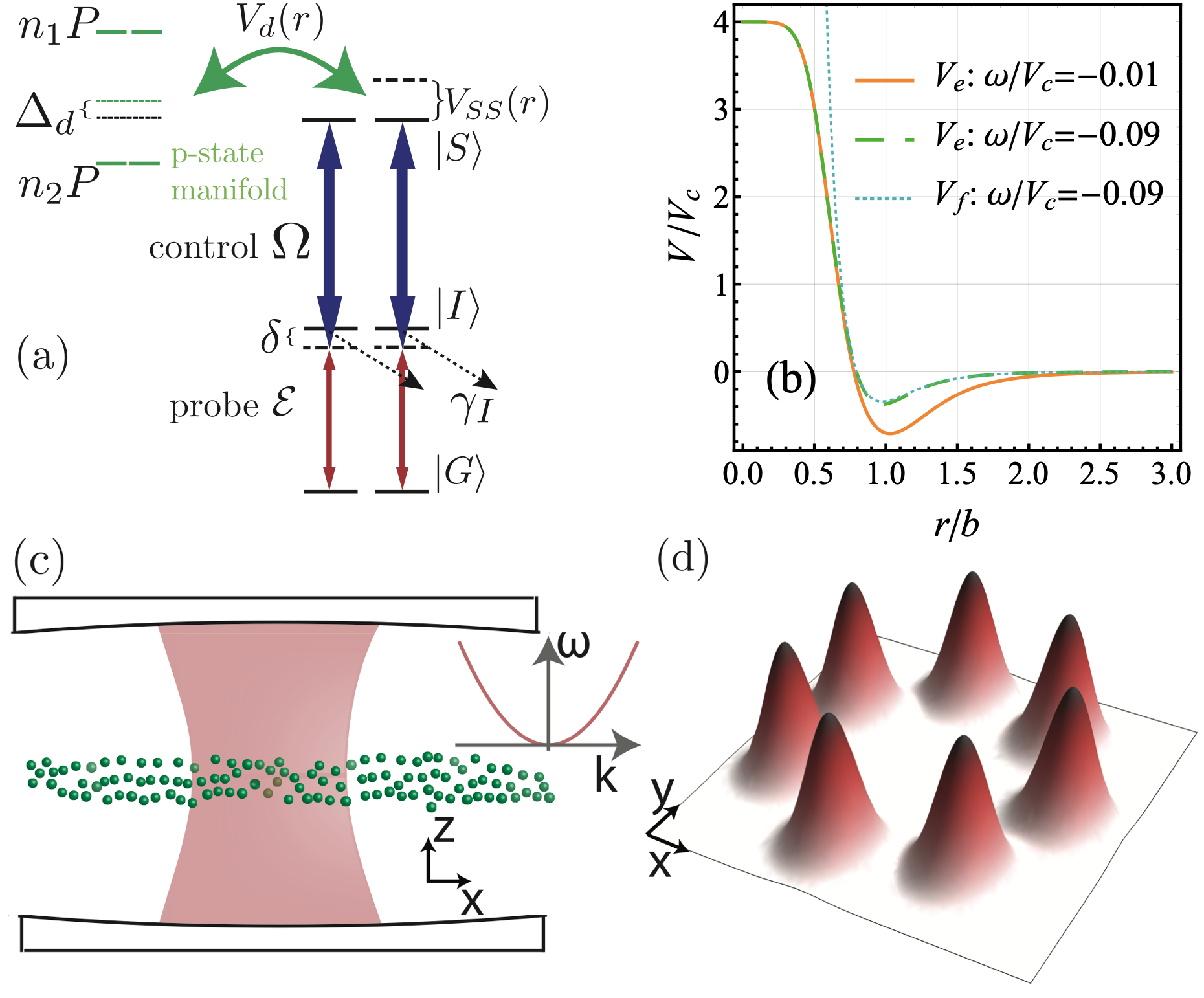}
\caption{
Using magnetic fields, we tune $\ket{SS}$ (characterized by principal quantum number $n$) close to the resonance (deviation from resonance denoted by $\Delta_d$) with the $P_1P_2$ state having $n_1$\,$=$\,$n$ and $n_2$\,$=$\,$n-1$. (b) This gives rise to the effective molecular potential $V_e$ plotted for $\Delta_d/V_c$\,$=$\,$0.03$ used also in \figref{self_consistent}(b). By working with $|V_{\rs min}|\ll 1/|\bar{\chi}|$, this potential around the local minimum is nearly equal to $V_f$ (i.e. is not modified by $\bar{\chi}$), see blue dotted vs green dashed curve. %
(c) A quasi-2D cloud of atoms placed in the center of a multi-mode cavity. %
The mode structure of the cavity gives an effective mass for free particles moving in 2D. The inset shows the resulting quadratic photonic dispersion relation.
For strong nonlinearities, %
this setup gives rise to few- and many-body self-bound clusters of light in 2D, e.g arranged on a ring, which is illustrated in (d) for seven photons.
}
\label{schematic}
\end{figure}\newparagraph 
\cmntc{``
A highly useful tool for controlling the interaction are field \Foerster resonances, where two dipole-coupled pair states are shifted into resonance by a dc [27] or microwave [28, 29] electric field. 
\Foerster resonances have been studied by observation of dipole blockade [30], line shape analysis [31], double-resonance spectroscopy [32], excitation statistics [33], and Ramsey spectroscopy [34, 35]. Recently, the anisotropic blockade on Forster resonance [36] and quasi- forbidden Forster resonances [37] have been observed and Forster resonances between different atomic species have been predicted [38]. For Rydberg-mediated single-photon transistors, the near-resonance in zero field for specific pair states has been used to enhance the transistor gain [19], while in experiments on Rydberg atom imaging [39, 40] an increase in Rydberg excitation hopping has been observed on resonance [41].''}
We achieve this using \Foerster resonances, which %
are a useful tool to control interactions between atoms ~\cite{Lukin2001,Afrousheh2004,Bohlouli-Zanjani2007,Vogt2006,Ryabtsev2010,Reinhard2008,Nipper2012,Ravets2015,Pelle2016,Beterov2015}. 
Application of these resonances to quantum optics with Rydberg polaritons was studied in the context of Rydberg atom imaging~\cite{Gunter2012,Olmos2011,Gunter2013} and an all-optical transistor~\cite{Gorniaczyk2016,Tiarks2014}.
We demonstrate that with an appropriate choice of states and couplings, we can achieve a \molecularlike potential between photons, which has a %
global minimum at a finite distance [see \figref{schematic}(b)].
We show the existence of bound states in 1D and 2D for two photons interacting via this molecular potential, and further %
discuss multi-photon self-bound clusters (molecules) of photons~[see \figref{schematic}(d)]. 

In the previous 
studies of shallow \cite{Firstenberg2013} and deep~\cite{Bienias2014} bound states, the photons interacted via a soft-core potential and therefore
preferred to overlap. %
  This %
  precluded the formation of more complex molecular-like structures
  that is possible in our proposal.
The many-photon clusters studied here resemble photonic crystalline features studied in 1D ~\cite{Otterbach2013,Chang2008}  and  2D~ \cite{Sommer2015}.
However, the latter proposals are based on  strong repulsion and therefore, without the external trapping potential, the crystals become \textit{unstable} in contrast to our work proposing \textit{self-bound} clusters.

One of the unconventional properties of \rEIT is 
strong three- and higher-body interactions between polaritons  \cite{Jachymski2016,Gullans2016,Kalinowski2020b}. 
These strong three-body interactions impact the energies of three-body bound states \cite{Liang2018}.
Here, we show that \Foerster resonances in combination with \rEIT lead to another source of many-body forces.
These additional forces  give rise to new phenomena. For example, it is energetically favorable to have three polaritons in a line, rather than in a triangular configuration. \spaceCutA{
Moreover, for four polaritons, by tuning the external parameters such as detuning and Rabi frequencies,
a transition can occur between two qualitatively-different configurations of the molecular ground state:
between a \lineconfiguration and the configuration in which two bound-pairs (dimers) are separated by a large distance. %
In the following, we present these results in detail.} %

\paragraph{System}Throughout, we focus on photons evolving in 1D and 2D multimode cavities~\cite{Sommer2015,Sommer2016,Schine2016,Jia2018,Clark2019b}, \figref{schematic}(c), described by the single-particle Hamiltonian~\cite{Sommer2015,Parigi2012,Stanojevic2013,Georgakopoulos2018b,Litinskaya2016,Grankin2014} ($\hbar$\,$=$\,$1$)
\begin{eqnarray}
H_1 &=&%
\int d\br
	\colthree{\hat{\E} }{\hat{I}}{\hat{S}}^\dag\!
	\left(\begin{array}{ccc}
	-i\kappa+\kinetic & g & 0\\
	g  & \Delta & \Omega\\
	0 & \Omega & -i\gamma_S\\
	\end{array}
	\right)\colthree{\hat{\E} }{\hat{I}}{\hat{S}}
\label{eq:Heit},
\end{eqnarray}
where $\hat{\E }$ is the field operator describing the photonic mode, whereas  $\hat{I}$ and $\hat{S}$  describe intermediate- and Rydberg-state collective spin excitations, respectively~\cite{Fleischhauer2000}.
$2 \kappa$ is the cavity loss rate, $\Delta$\,$=$\,$\delta-i\gamma_I$ is the complex single-photon detuning, $2\gamma_I$ is the atomic intermediate state decay rate, $2\gamma_S$ is the Rydberg level decay rate, $g$ is the single-photon coupling, and $\Omega$ is the Rabi frequency of the control drive. 
The kinetic energy of photons is described in 1D and 2D via $\kinetic$\,$=$\,$-\frac{\nabla^2}{2m_{\rs ph}}$, where $m_{\rs ph}$ is the photon mass defined by the cavity parameters.
Note that our approach can be easily generalized to a 1D free-space geometry, which is discussed below. 
The Hamiltonian~ in Eq.\ \ref{eq:Heit} can be diagonalized and leads to two bright and one dark polariton branches~\cite{Bienias2014}.  
Well within the EIT window~\cite{Petrosyan2011} and in the limit of  $\Omega\ll g$ (assumed throughout), the dark-state polariton $\hat{D}$  takes the form $\hat{D}\sim \hat{S}-\frac{\Omega}{g}\hat{\E} $.  %
To leading order, the dark-state polariton losses are $\frac{\Omega^2}{g^2}\kappa+\gamma_S$ and are negligible for the evolution times considered in this Letter. %
For simplicity we shall assume $|\delta|\gg \gamma_I$, and therefore neglect the imaginary part of $\Delta$.
The dispersion of $\hat{D}$ is inherited from the photonic component and therefore described via an enhanced mass equal to
$m$\,$=$\,$\frac{g^2}{\Omega^2}m_{\rs ph}$.

The interactions for conventional dark-state polaritons are inherited from the van der Waals (vdW) interactions between Rydberg states~\cite{Mohapatra2008,Friedler2005} described by the quartic term proportional to $\hat{S}^\dagger(\br) \hat{S}^\dagger(\br') V_{SS}(\br-\br')\hat{S}(\br')\hat{S}(\br)$.
However, close to the \Foerster resonance, the physics becomes more subtle because at least two strongly-interacting pairs of states are involved. 
To build intuition, we first study the two-body problem\spaceCutA{, and afterward the multi body problems}.

\paragraph{Effective \molecularlike potential}
In the past, \Foerster resonances were used in \rEIT transistor experiments~\cite{Gorniaczyk2016,Tiarks2014} which used two $S$-states with different principal quantum numbers for the gate and source photons. 
Here, we are interested in few- and many-body physics and therefore 
use a single $nS$-state~\cite{foot2}.
In this case, there is no true \Foerster resonance at zero external fields \cite{Walker2008}, but there is an approximate one $n S + nS \ra n P+ (n-1) P$. 
 We consider $J$\,$=$\,$1/2,m_J$\,$=$\,$1/2$ $S$-states and $J$\,$=$\,$3/2,m_J$\,$=$\,$3/2$ $P$-states and tune them to near resonance~\cite{Gorniaczyk2016} [see \figref{schematic}(a)] using a strong magnetic field 
(defining the quantization axis) %
perpendicular to the atomic cloud \cite{foot3}.
%
%
%
%
%

%
Under these conditions, there are three relevant pairs of Rydberg states $\{ SS,P_1P_2,P_2P_1 \}$ with interactions between them described by
\beqa
\left(
\begin{array}{ccc}
 V_{SS} & V_d & V_d \\
 V^*_d & V_{PP}+\Delta_d & V_{PP,\text{off}} \\
 V^*_d & V_{PP,\text{off}} & V_{PP} +\Delta_d\\
\end{array}
\right),
\label{HintForster}
\eeqa
where $\Delta_d$\,$=$\,$E_{P_1}+E_{P_2}-2E_S$ is the \Foerster defect and $V_d$\,$=$\,$C_{3}e^{i2\phi_{12}}/r^3$ is a dipolar interaction with the polar angle $\phi_{12}$ describing the direction of the relative distance  $\br$\,$=$\,${\bf r}_1-{\bf r}_2$ between the first and second excitation, with $r$\,$=$\,$|\br|$. In general, $|V_d|$ could have an additional azimuthal-angle dependence, which  is not present for the 1D and 2D geometries considered here. $V_{SS}$\,$=$\,$C_{SS}/r^6,V_{PP}$\,$=$\,$C_{PP}/r^6$ are diagonal vdW interactions \cite{foot4}
, whereas 
 $V_{PP,{\rs off}}$\,$=$\,$C_{PP,{\rs off}}/r^6$  is the off-diagonal vdW interaction between $P_1P_2$ and $P_2P_1$.
The conventional \rEIT two-body problem can be described using a set of nine coupled Maxwell-Bloch equations~\cite{Gorshkov2011,Peyronel2012,Bienias2016}  for $XY$ components of the two-body wavefunction, where  $X,Y\in\{\E ,I,S\}$.
Importantly, the $P_2P_1$ and  $P_1P_2$ components are coupled to the conventional equations~\cite{Peyronel2012,supplementCluster} only via dipolar interactions $V_d$. 
This enables us to eliminate the $P_2P_1$ and  $P_1P_2$ components  (see supplement~\cite{supplementCluster}) and leads to the standard equations of motion but where $V_{SS}$ is replaced by 
\beq
V_f(r)=
\frac{C_{SS}}{r^6}-
\frac{2
\left(\frac{C_{3}}{r^3}\right)^2
}{\Delta_d+ \frac{C_{PP}+ C_{PP,{\rs off}}}{r^6}-\omega} 
\label{eq:Vf}
\eeq
with $\omega$ the total energy of the pair of polaritons.
This potential [see Fig.~\ref{schematic}(b)] can have a local minimum which intuitively comes from the interplay of the diagonal interactions $\sim 1/r^6$ and the off-diagonal couplings $\sim 1/r^3$: 
The latter terms dominate at large separation causing the potential curve $SS$ to be attractive, 
 whereas at short distances the vdW interaction dominates and the potential is repulsive.
This is in contrast to other molecular potentials~\cite{Hollerith2019,Petrosyan2014} %
arising from the avoided crossings between potential curves.
Based on $V_f$, using the approach developed in Ref.~\cite{Bienias2014}, 
we arrive at %
the soft-core effective potential between polaritons 
\be \label{eqn:Ve}
V_e({r}) =  \frac{  V_f({r})}{1- \bar{\chi} V_f({r})}, %
\ee
where
$\bar{\chi}$\,$=$\,$\frac{\Delta}{2\Omega^2} - \frac{1}{2\Delta}$ for the regime considered below.
In contrast with conventional Rydberg-EIT, both the strength and shape of $V_f$ can be tuned using $\Delta_d$ and the choice of principal quantum numbers.
In general, the depth $V_{\rs min}$ of the potential $V_e$ can be as large as its height equal to $-1/\bar{\chi}$. However, by assuming henceforth a shallow $V_e$ such that $V_{\rs min} \ll \omega_c\equiv 1/|\bar{\chi}|$, we can: 
(i) neglect dependence %
of $\bar{\chi}$ on $\omega$ because $\omega\sim V_{\rs min}$~\cite{Bienias2014};
(ii) neglect the scattering to bright polaritons for %
a small center-of-mass momentum $K\ll k_c\equiv\frac{g^2}{c\Omega^2}\omega_c$~\cite{Bienias2014} assumed henceforth; %
(iii) neglect blockade-induced three-body interactions~\cite{Jachymski2016,Gullans2016}. %
\paragraph{Two-body problem} 
The two-body problem can be described using the wavefunction $\varphi(\bf{r})$ depending on the relative distance $\bf{r}$.  $\varphi$ describes two %
dark-state polaritons, is proportional to $\mathcal{E E}\sim \mathcal{E S}+\mathcal{SE}$,  and is the solution of the  effective Schroedinger equation \cite{Bienias2014}
\be \label{eqn:Schroedinger}
\omega\varphi({\bf r}) =  \bigg[ - \frac{\nabla^2}{ m} +
V_e({ r},\Delta_d,\omega) \bigg] \varphi(\bm{r}). 
\ee
As discussed, 
we can neglect dependence of $\bar{\chi}$ %
on  $\omega$,
however, there is still dependence of $V_f$ on $\omega$, see \eqref{eq:Vf}, which we take into account in the numerics.
The local minimum of $V_e$ exists for 
$C_{SS} \left(\omega -\Delta _d\right)+2C_3^2>0$. 
Considering $|\omega|\ll\Delta_d$ enables us to define the characteristic energy $\nu_c$\,$=$\,$2C_3^2/C_{SS}$  quantifying the range of $\Delta_d$ for which a bound state could exist.
In addition, we define (for details see supplement) the characteristic length scale $b$\,$=$\,$\left({ \left(\sqrt{2}+1\right) C_{PP} C_{SS}}/{C_3^2}\right)^{1/6}$ for the position of the local minimum, and
$V_c$\,$=$\,${2C_3^2 }/{C_{PP}}$ quantifying the depth of the potential. %

\begin{figure}[h!]
\includegraphics[width= .99\columnwidth]{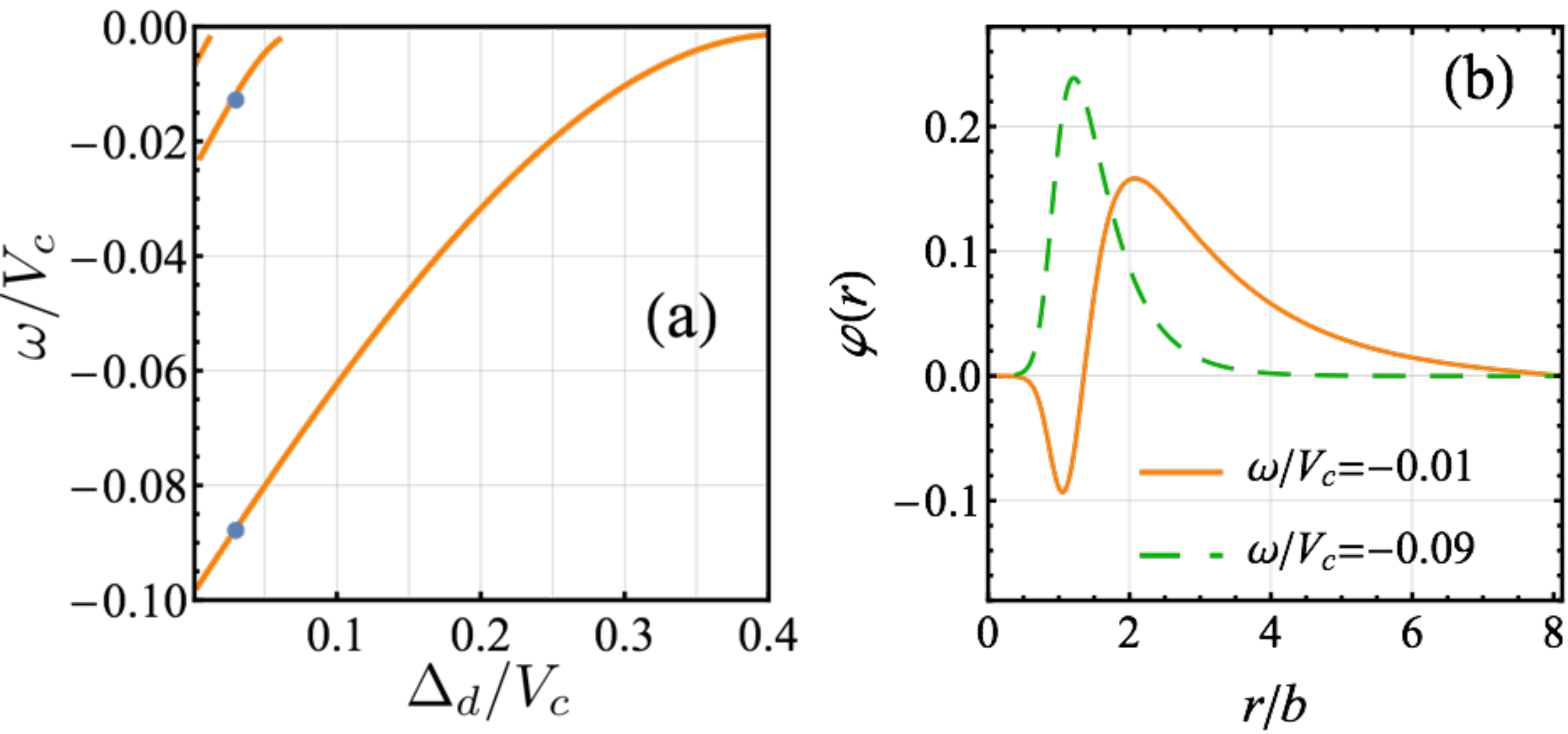}
%
%
%
\caption{
Results for $^{87}$Rb, $n$\,$=$\,$120,$ $n_1$\,$=$\,$120,$ $n_2$\,$=$\,$119$, $\Omega<|\Delta|$, $V_cm\,b^2$\,$=$\,$40$, 
and $\omega_c$\,$=$\,$4 V_c$, which is achieved by an appropriate choice of $\Omega$. %
(a)    Bound-state energies as a function of $\Delta_d$ in units of 
$V_c$. 
   (b) The wavefunctions for two lowest bound states with $\Delta_d/V_c$\,$=$\,$0.03$ (blue dots in (a)).  %
}
\label{self_consistent}
\end{figure}

\spaceCutA{\paragraph{Self-consistent solution for the two-body bound state}}
Next, we self-consistently find the solutions of \eqref{eqn:Schroedinger} for different $\Delta_d$.%
\cmntc{Since $g\gg\Omega$ we can set $\alpha$\,$ \approx $\,$ 1$.}
\new{In \figref{self_consistent}, we show solutions for the 1D limit of \eqref{eqn:Schroedinger} for $V_cm\,b^2$\,$=$\,$40$, corresponding to $\OD_b\, {\gamma_I}/{\delta}$\,$=$\,$0.9$, where OD$_{b}\equiv\OD \frac {b}{ L}$ is an optical depth \OD per $b$.}
The smaller the ratio $\Delta_d/V_c$ is, the deeper $V_{f}$ is and, therefore, the second (and even third) bound state can be seen in~\figref{self_consistent}(a). 
\spaceCut{The lowest bound state (green in ~\figref{self_consistent}(b)) has a width smaller than the first excited bound state (orange), the latter having a single node around the local minimum of $V_e$ at $r$\,$ \approx $\,$ b$, \figref{schematic}(b).} 
Both wavefunctions are strongly suppressed at short distances due to the strong repulsion for small~$r$.

\paragraph{Three and more photons}
For conventional few-body problems, it is usually a good approximation to assume that each pair of bodies interacts via a two-body potential.
However, Rydberg polaritons are an unconventional platform enabling strong many-body interactions, as was shown for a soft-core potential $V_e$ in Refs~\cite{Jachymski2016,Gullans2016,Liang2018}. 
In this Letter, we can neglect these higher-body interactions because states of interest are largely supported outside the repulsive core of the potential, 
and therefore, the three-body forces are strongly suppressed for $|\bar{\chi}V_{\rs min}|\ll 1$, %
 Refs~\cite{Jachymski2016,Gullans2016,Kalinowski2020b}. 
However, we show that \Foerster resonances in combination with \rEIT lead to another source of many-body forces.
\spaceCutA{This is easiest to see in the specific example of the three-body problem.}

\spaceCutA{For the sake of clarity of presentation, with the full description presented in the Supplement~\cite{supplementCluster}, we further simplify our problem:}
Even though the $SS$ channel is on resonance with two channels $P_1P_2$ and $P_2P_1$, the majority of the three-body physics can be well-described by a single effective channel $PP\sim P_1P_2 + P_2P_1$~\cite{supplementCluster}%
(note that all the numerics are performed without this approximation). 
The dipolar interaction between states $\{SSS,SPP,{PSP},{PPS}\}$ takes the form
\beq
\left(
\begin{array}{cccc}
V_S & V_{d,23} & V_{d,13} & V_{d,12} \\
 V^*_{d,23} & V_{{P,1}}+\Delta _d &W_{12} & W_{13} \\
 V^*_{d,13} & W_{12} & V_{{P,2}}+\Delta _d &W_{23} \\
 V^*_{d,12} & W_{13} & W_{23} & V_{{P,3}}+\Delta _d \\
\end{array}
\right),
\label{H_dark_states}
\eeq
where $V_{P,i}$\,$=$\,$V_{PP}(r_j-r_k) + V_{SP}(r_i-r_j) + V_{SP}(r_i-r_k)$ with $i\neq j,k$ and $j < k$
describes all vdW interactions between involved Rydbergs; $V_S$ is  a sum of vdW interactions between all polaritons in $S$ state. $V_{d,ij}$\,$=$\,$\sqrt{2}C_3e^{i2\phi_{ij}}/|\br_i-\br_j|^3$ %
 is the effective dipole interaction between $SS$ and $PP$. %
Analogously, $W_{ij}(r)$\,$=$\,$-\frac{1}{3}C_3/|\br_i-\br_j|^3$ describes the dipole interaction between $SP$ and $PS$. 
Without off-diagonal $W$ terms, we could eliminate all components containing $P$-states. However, due to these exchange terms, this is no longer possible, which is one of the reasons for the strong N-body forces.

\paragraph{Low-energy regime\label{sec:Dark}}
The low energy assumption, $\kinetic_i\ll |V_{\rs min}|$ (where $\kinetic_i$ is the kinetic energy of the $i$th polariton), together with the already made assumption that $|V_{\rs min}|\ll \omega_c$, ensures that the dipolar interactions modify the internal composition of the dark states only weakly~\cite{Otterbach2013}. %
Therefore, we can neglect the blockade effects on the effective interaction $V_e$ and 
dark-state polaritons. %
In the slow-light regime of $g\gg\Omega$, dark states $D$ have a negligible contribution from $\E $ and $I$ and mostly consist of Rydberg states.
Hence, the dipolar Hamiltonian \eqref{H_dark_states} maps directly onto dark-state polaritons $D$ and collective excitations $P$.
That is, the full Hamiltonian describing the evolution of the three polaritons in the $\{\Dpol\Dpol\Dpol,\Dpol\Fpol\Fpol,{\Fpol\Dpol\Fpol},{\Fpol\Fpol\Dpol}\}$ basis is a sum of \eqref{H_dark_states} with kinetic terms $\{ \kinetic_1 +\kinetic_2+\kinetic_3,\kinetic_1,\kinetic_2,\kinetic_3\}$ on the diagonals.

\paragraph{Few-body bound states in the large-mass limit}
To give additional insights into the role of few-body interactions in the many-body problem, in the following we neglect kinetic energy all together.
This requires $V_c\gg 1/{m } b^2 $ and therefore OD$_{b}\gg \frac{\delta}{\gamma_I}$
%
%
%
%
%
.
\begin{figure}[htbp]
\begin{center}
\includegraphics[width=.999 \columnwidth]{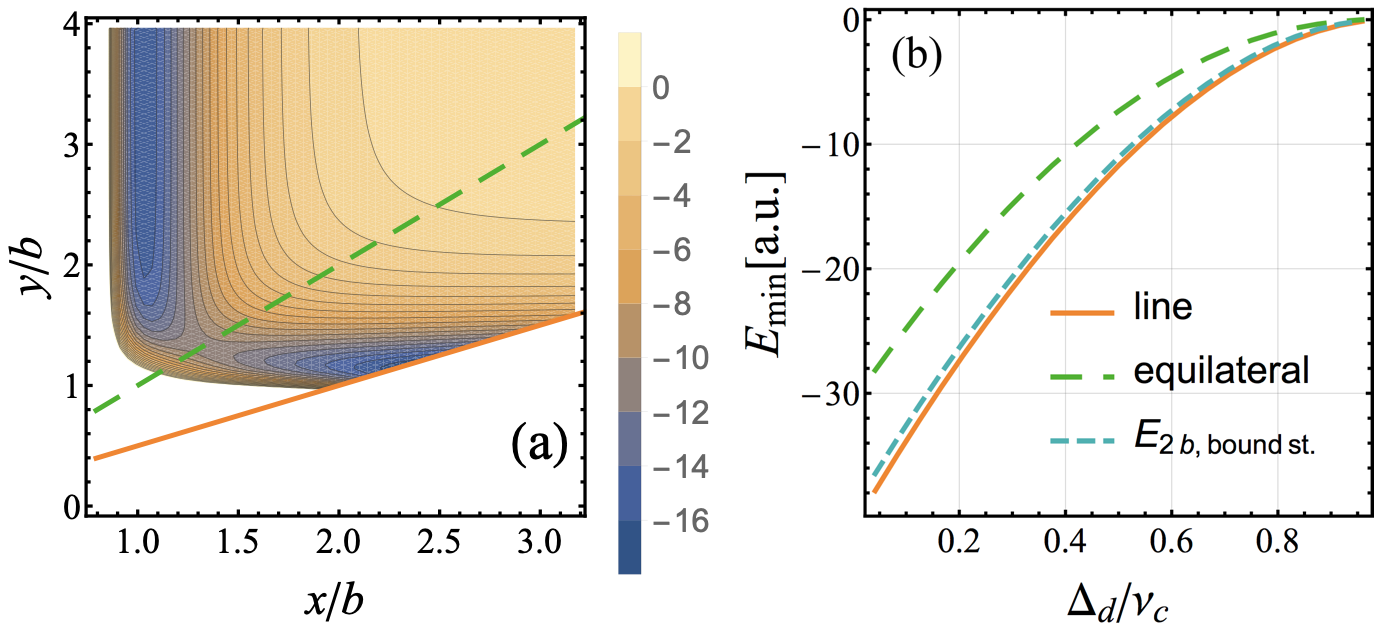}
\caption{Self-consistent solution of \eqref{H_dark_states} describing polaritons in the large-mass limit.
(a) Three-body problem in the isosceles triangular configuration with edge lengths $y,y,x$.  %
(b) Lowest energy  as a function of $\Delta_d$ for line, regular-polygon, and dimer configurations for three  bodies. 
Results are for $^{87}$Rb with $n_{1},n_2,n$ as in \figref{self_consistent}. Additionally, (a) is for $\Delta_d/\nu_c$\,$=$\,$0.4$. 
}
\label{fig:selfconsistent}
\end{center}
\end{figure}
Next, we numerically solve for the eigenstates of the two-channel version~\cite{supplementCluster} of the Hamiltonian \eqref{H_dark_states} as a function of separations $r_{ij}$ in a 2D geometry.
We find that it is preferable to have three polaritons in a line rather than in an equilateral-triangle configuration.
Moreover, the configuration in which one photon is away from the dimer has lower energy than the equilateral-triangle configuration. 
This is demonstrated in \figref{fig:selfconsistent}(a-b) where (a) shows
total energies for the polaritons being at the corners of an isosceles triangle
and (b) shows that  %
regardless of the value of $\Delta_d/\nu_c$, the line-configuration has the lowest energy~\cite{foot5}. 
\paragraph{Intuition behind the N-body force}
For the three-body problem, even approximate analytical expressions for eigenstates of \eqref{H_dark_states}  are lengthy %
for arbitrary separations.
Therefore, we use an equilateral-triangle configuration parametrized by edge length $r$ to obtain more intuition on the three-body forces.   
The energy $E$ being the lowest eigenvalue of \eqref{H_dark_states} as a function of the separation $r$, for $E\ll\Delta_d$, takes the form 
\beqa
E=3\left(\frac{ C_{SS}}{r^6}-\frac{2 \left(\frac{\text{C}_3}{r^3}\right){}^2}{\Delta_d+
\frac{C_{PP}+2 C_{SP}}{r^6}- W(r)}\right). %
\label{intuition}
\eeqa
From comparison of this expression with \eqref{eq:Vf}, we see
that the denominator has two additional terms~\cite{foot6}: 
 (i) the vdW energy shift $2C_{SP}/r^6$, and (ii)
the shift due to the off-diagonal dipolar interactions $W(r)<0$~\cite{foot7}. 
Both lead to the suppression of $E$%
, resulting in strong three-body repulsion which prevents configurations with three particles closely spaced. This gives rise to the novel ground state geometries presented  in~\figref{fig:selfconsistent}.
\remove{
\paragraph{ Four-body problem} The four-body problem features additional exotic phenomena due to the strong four-body interactions.
For $^{87}$Rb, the ground state is a configuration consisting of two far-separated dimers~\cite{foot8}, 
see \figref{fig:selfconsistent}(c). %
However for $^{133}$Cs, which has $C_{SP}/C_{PP}$\,$ \approx $\,$ 0.6$ (compared with $C_{SP}/C_{PP}$\,$ \approx $\,$ 1.4$ for $^{87}$Rb) and therefore weaker multi-body forces, 
the ground state configuration depends on $\Delta_d$, \figref{fig:selfconsistent}(d): %
the ground state is a linear configuration
for $\Delta_d/\nu_c\geq 0.3$ and two far-separated dimers for $\Delta_d/\nu_c\leq 0.3$ \cite{foot9}.}
 %
%
%
%

%
{
\paragraph{Multiple-body problem}
  \begin{figure}[htbp]
\begin{center}
 %
%
%
\includegraphics[width=.99 \columnwidth]{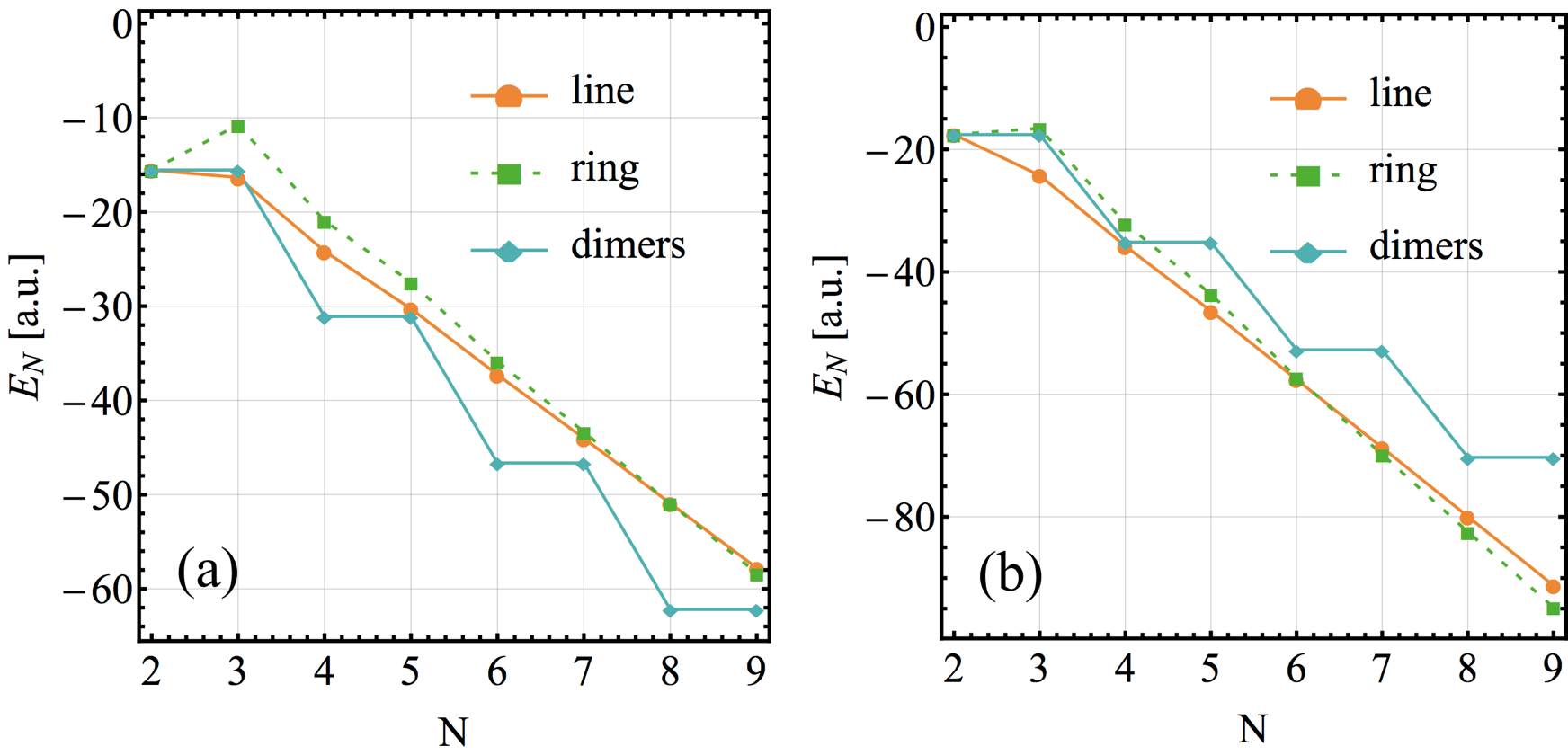}
\caption{
The lowest energy for on-the-line and on-the-ring configurations for  $\Delta_d/\nu_c$\,$=$\,$0.4$ for (a) Rb  and (b) Cs. 
We see that strong N-body forces lead to different geometry of the ground state depending on $N$ and the atomic species. 
}
\label{fig:fewBody}
\end{center}
\end{figure}
From \figref{fig:fewBody}(a), we see that the few-body forces lead to 
an effect in which many photons prefer to be arranged as independent dimers. 
For Cs [see \figref{fig:fewBody}(b)]
at large enough $N$ (which depends on $\Delta_d/\nu_q$, and for $\Delta_d/\nu_q$\,$=$\,$0.4$ happens for $N\geq 7$), a regular polygon (ring) is the ground state rather than a \lineconfiguration [see~\figref{schematic}(d)]. Intuitively, the additional two-body attractive bond for the ring arrangement wins over the additional repulsive many-body forces present in this configuration.

\paragraph{Experimental realization} 
Photons in a multi-mode cavity~\cite{Sommer2015} enable us to tune the polariton's mass so that $m V_e$ is repulsive at short distances, has local minima at finite distance, and is free of potentially lossy %
divergences~\cite{Maghrebi2015c,Bienias2014,Gullans2017}. %
In general, for multi-mode cavities, the generation of the mass is intertwined with the presence of the trapping potential.
However, for a near-planar cavity (defined as $R\gg L$, where $R$ is mirror curvature and $L$ distance between mirrors) we have $m_{\rs ph}$\,$=$\,$\frac{%
k}{c}$ and the trapping frequency $\omega_{\rs tr}$\,$=$\,$\frac{c}{\sqrt{2 LR}} $. Therefore, the trapping vanishes with increasing $LR$. %

Note that our scheme also works in a free-space quasi-1D geometry~\cite{Peyronel2012} for a magnetic field perpendicular to the propagation direction and the transverse mass much greater than the longitudinal one~\cite{Gullans2017}.
Then, by working in the regime $\Omega>|\Delta|$, we can achieve a divergence-free potential that is repulsive at short distances~\cite{Maghrebi2015c,Bienias2014}.

\new{
To prepare the ground state on small systems we envision  a spectroscopic post-selection approach whereby  a weak product coherent state  wavefunction is input with the mode frequencies chosen to add up to the energy of the target ground state. The mode functions of the photons  can be further chosen to maximize the ground-state overlap.  State preparation is then possible through post-selection on the total photon number.  The weak input condition ensures that the target manifold is not spoiled by dissipation from higher-photon number manifolds.  For larger systems, more efficient preparation schemes become necessary that are still insensitive to dissipation.  We imagine using dissipative Rydberg blockade~\cite{Peyronel2012,Dudin2012} to prepare a product state of many single photons as the starting state for adiabatic transfer to the ground state.  This method is robust to dissipation for ground state gaps larger than the  dissipation rate. Once the state is prepared, the measured multi-body 2D correlation functions can be    postprocessed~\cite{Schauss2012} to prevent the rotational symmetry and shot-to-shot variations in measurement outcomes   from smearing out spatial patterns. We leave a more complete and detailed analysis of preparation and detection for future work.}

\paragraph{Outlook}
In this work, we concentrated on the strongly interacting regime in 1D and 2D. 
Another direction is a study of the 3D interacting regime of photons copropagating in free space in the presence of the molecular potential in the transverse directions~\cite{Sevincli2011,Gullans2017}.
Note that our analysis suggests that the strong few-body forces can also be observed in experiments with ultracold Rydberg atoms alone~\cite{Schauss2012b,Zeiher2017} \new{rather than Rydberg-polaritons.}
This can be done in a 2D pancake geometry with or without an additional optical lattice potential. It is an especially promising direction in the light of recent work on the observation of Rydberg macrodimers~\cite{Hollerith2019} with $P$ states close to \Foerster resonances.
}

\begin{acknowledgments}\paragraph{Acknowledgments}
We thank H. P. Buechler, S. Hofferberth, I. Lesanovsky, J. Young, Y. Wang, and S. Weber 
for insightful discussion.  
P.B., M.K., A.C, D.O.-H, S.L.R., J.V.P., and A.V.G. acknowledge support from the United States Army Research Lab’s Center for Distributed Quantum Information (CDQI) at the University of Maryland and the Army Research Lab, and support from the National Science Foundation Physics Frontier Center at the Joint Quantum Institute (Grant No. PHY1430094).  P.B., M.K., and A.V.G. additionally acknowledge support from AFOSR, ARO MURI, DoE ASCR Quantum Testbed Pathfinder program (award No. DE-SC0019040), DoE BES Materials and Chemical Sciences Research for Quantum Information Science program (award No. DE-SC0019449), DoE ASCR FAR-QC (award No. DE-SC0020312), and NSF PFCQC program. %
M.K. acknowledges financial support from the Foundation for Polish Science within the First Team program co-financed by the European Union under the European Regional Development Fund.
\end{acknowledgments}

%
%
 
\bibliographystyle{bernd} 
\bibliography{library,additional}

%
%

\clearpage
\begin{widetext}
\begin{center}
{\Large \centering Supplemental material}
\end{center}

\input{supplement_text.tex}
\clearpage
\end{widetext}

\end{document}

%% file: supplement_text.tex

Here, we present the derivation of effective interactions between polaritons propagating through Rydberg media close to the \Foerster resonance (sec.~\ref{twoPolar}), derivation of the characteristic energy and length scales in the two-body problem (sec.~\ref{units}), derivation of the single-channel description used to give intuition behind three-body forces (sec.~\ref{effectiveChannel}), and self-consistent solution of the four body problem (sec.~\ref{fourBody}).

\section{Two photons propagating through Rydberg media close to the \Foerster resonance \label{twoPolar}}
Here, we give more details related to the effective interactions between Rydberg states described by \eqref{eq:Vf} in the main text.
Our model system is a one-dimensional gas of atoms whose electronic levels are given in Fig.~1(a) in the main text. 
Following Ref.~\cite{Gorshkov2011,Peyronel2012,Bienias2014,Gorniaczyk2016}, we introduce operators $\hat{I}^\dagger(z)$ and $\hat{S}^\dagger(z)$ which generate the atomic excitations into the  $\ket{I}$ and $\ket{S}$ states, respectively, at position $z$.
In addition, comparing to Ref.~\cite{Gorshkov2011,Peyronel2012,Bienias2014,Bienias2016a,Gorniaczyk2016} we include a more complex atomic level structure of the source and the gate excitations %
by defining $\hat{\P}_1^\dagger(z)$ and $\hat{\P}_2^\dagger(z)$ which create excitations into $\ket{P_1}$ and $\ket{P_2}$ states, respectively.
All the operators $\hat{O}(z)\in\{\hat{\E }(z)\,,\hat{I}(z)\,,\hat{S}(z)\,,\hat{\P_1}(z)\,,\hat{\P_2}(z)$ are bosonic and satisfy the equal time commutation relation, $[\hat{O}(z),\hat{O}^{\dagger}(z')]=\delta(z-z')$. %

The microscopic Hamiltonian  describing the propagation consists of three parts: $\hat{H}=\hat{H}_\text{p}+\hat{H}_\text{ap}+\hat{H}_\text{int}$.
For the sake of simplicity we show the derivation for the 1D massive photons, which straightforwardly applies to the free-space photons within the center of mass frame, and generalizes to a 2D cavity.
The first term describes the photon evolution in the medium and is defined as
\begin{equation}
\hat{H}_{\text{p}}= -\frac{1}{2m_{\rs ph}}\int dz\hat{\E }^{\dagger}(z)\partial^2_z\hat{\E }(z), 
\end{equation}  
 with the mass defined by the cavity geometry. The atom-photon coupling is described by 
 \begin{eqnarray}
\hat{H}_\text{ap}=\int dz &\bigg{[} 
& g\hat{\E }(z)\hat{I}^{\dagger}(z)+\Omega\hat{S}^{\dagger}(z)\hat{I}(z) 
+ g\hat{{I}}(z)\hat{\e}^{\dagger}(z)+\Omega\hat{I}^{\dagger}(z)\hat{S}(z)  + %
{\Delta}{} \hat{I}^{\dagger}(z)\hat{I}(z)  
\bigg{]},
 \end{eqnarray}
  where $g$ is the collective coupling of the photons to the matter, %
  and for the sake of brevity  we drop the decay rates $\gamma_S$, $\gamma_{P_{1}}$ and $\gamma_{P_{2}}$. 
The interaction between Rydberg levels is described by 
\beqa
	\hat{H}_\text{int}=
\frac{1}{2}\int dz' \int dz
	\colthree{\hat{S}\hat{S}}{\hat{P_1}\hat{P_2}}{\hat{P_2}\hat{P_1}}^\dag\!
\left(
\begin{array}{ccc}
 V_{SS} & V_d & V_d \\
 V^*_d & V_{PP}+\Delta_d & V_{PP,\text{off}} \\
 V^*_d & V_{PP,\text{off}} & V_{PP} +\Delta_d\\
\end{array}
\right)
		\colthree{\hat{S}\hat{S}}{\hat{P_1}\hat{P_2}}{\hat{P_2}\hat{P_1}},
\label{HintForster}
\eeqa
where the notation was explained in the main text.
The Schroedinger equation has the form
\beqa
i\hbar\partial_t\ket{\psi(t)}=\hat{H} \ket{\psi(t)},
\label{Schroedinger}
\eeqa
 with the two-excitation wavefunction having the form~\cite{Peyronel2012,Bienias2016a}
\beqa 
\ket{\psi(t)}&=& \integral{z}\integral{\z2} {\bigg [} \left. 
\frac{1}{2}     {\E\E}(z,\z2,t)  \psiOp{\E}^\dagger(z)   \psiOpDr{\E}(\z2)+ \frac{1}{2}    {\pSt\pSt}(z,\z2,t)  \psiOp{\pSt}^\dagger(z)   \psiOpDr{\pSt}(\z2)+
\frac{1}{2}        {\rSt\rSt}(z,\z2,t)  \psiOp{\rSt}^\dagger(z)   \psiOpDr{\rSt}(\z2) \right.\\
&+&\left. 
{ {\E}\pSt}(z,\z2,t)  \psiOp{\E}^\dagger(z)   \psiOpDr{\pSt}(\z2)+
 { \E\rSt}(z,\z2,t)  \psiOp{\E}^\dagger(z)   \psiOpDr{\rSt}(\z2) + 
 {\pSt\rSt}(z,\z2,t)  \psiOp{\pSt}^\dagger(z)   \psiOpDr{\rSt}(\z2)+
{P_1P_2}(z,\z2,t)  \psiOp{P}^\dagger_1(z)   \psiOpDr{P_2}(\z2) \nn
\right. {\bigg]}
        \ket{0}
. \eeqa
The Schroedinger equation~\cite{Peyronel2012} in the frequency space reduces to
\begin{align}\label{eq:me}
\omega \E\E(z,\z2)&= -\frac{1}{2m_{\rs ph}} \left(\partial^2_z +\partial^2_{\z2}\right) \E\E(z,\z2)+
g(\E \I(z,\z2) + \E\I (\z2,z) ),\\
\omega \E \I(z,\z2)&=\left(-\frac{1}{2m_{\rs ph}} \partial^2_z+\Delta\right)\E \I(z,\z2)+g \I\I(z,\z2)+\Omega \E S(z,\z2),\\
\omega \E S(z,\z2)&=\left(-\frac{1}{2m_{\rs ph}} \partial^2_z+\Delta\right)\E S(z,\z2)+g \I S(z,\z2)+\Omega  \E\I(z,\z2),\\
\omega \I\I(z,\z2)&=2\Delta\I\I(z,z')+
g(\E\I(z,\z2)+\E\I (\z2,z))+\Omega (\I S(z,\z2)+\I S(\z2,z)),\\
\omega \I S(z,\z2)&=\Delta \I S(z,\z2)+
g\E S(z,\z2)+\Omega SS(z,\z2),\\
\omega SS(z,\z2)&=
\Omega (\I S(z,\z2)+ \I S (\z2,z) ) + 
V_{SS}(z-\z2)SS(z,\z2)+V_d(z-\z2)\left(\P_1\P_2(z,\z2)+\P_1\P_2(\z2,z)\right),\\
\omega\P_1\P_2(z,\z2)&=
V^*_d(z-\z2)SS(z,\z2)+V_{PP}(z-\z2)\P_1\P_2(z,\z2)+\D_d\P_1\P_2(z,\z2)+V_{PP,{\rs off}}(z-\z2)\P_1\P_2(\z2,z),%
\label{eq:s_me}
\end{align}
where only the two last equations differ from the conventional one~\cite{Gorshkov2011,Peyronel2012,Bienias2014,Gorniaczyk2016}.

\new{
Next, we eliminate the $\P_1\P_2$ component
\begin{align}
\P_1\P_2(z,\z2)&=\frac{V^*_d(z-\z2)}{\omega  -V_{PP}(z-\z2)-\D_d-V_{PP,{\rs off}}(z-\z2)}
SS(z,\z2),%
\end{align}
 which is not coupled by the laser field directly to photons. This leads to}
\begin{align}%
\omega SS(z,\z2)&=\left(V_{SS}(z-\z2)-\frac{2V_d(z-\z2)^2}{\Delta_d+V_{PP}(z-\z2)+V_{ PP,{\rs off}}(z-\z2)-\omega}\right)SS(z,\z2)+\Omega (\I S(z,\z2)+\I S (\z2,z)). %
\label{eq:s_me}
\end{align}
Note that $S(z,z')=S(z',z)$.
We see from Eq.\ (\ref{eq:s_me}) that the effective interaction between Rydberg states takes the form shown in Eq.~(3) 
in the main text. 

%

\section{The characteristic energy and lengths scales in the two body problem \label{units}}
Let us next comment  in more detail on the form of the $V_f(r)$ given by \eqref{eq:Vf} in the main text.
Since $|V_f|\ll \omega_c$, the depth of  $V_e$ is nearly equal to the depth of $V_f$ in the considered regime, %
and is given by
$
V_{\rs min}=-{(\sqrt{2}C_3 -\sqrt{C_{SS} \left(\Delta _d-\omega \right)}
)^2}/{C_{PP}}$; note that we consider states for which $C_{PP},C_{SS}$>0. 
The minimum of the potential occurs at the relative distance given by
$ 
r^6={C_{PP} \sqrt{C_{SS}}}/$$\left({\sqrt{2}C_3 \sqrt{\Delta _d-\omega}-\sqrt{C_{SS}} (\Delta _d-\omega)}\right)
\label{distanceForMin},
$ 
which leads to a characteristic length scale $b=\left({ \left(\sqrt{2}+1\right) C_{PP} C_{SS}}/{C_3^2}\right)^{1/6}$, by taking $\Delta_d=\nu_c/2$ with $\nu_c=2C_3^2/C_{SS}$.
The potential's local minimum exists for 
$C_{SS} \left(\omega -\Delta _d\right)+2C_3^2>0$. %
For $\omega=0$ %
and $\epsilon=\Delta_d/\nu_c <1$   we have $V_{\rs min} = -{2C_3^2 \left(\sqrt{\epsilon }-1\right)^2}/{C_{PP}}$. Therefore, we define $V_c={2C_3^2 }/{C_{PP}}$, which together with $\nu_c$  is used as a characteristic energy scale in our results.

\section{Comparison of single-channel $PP$ vs double-channel $P_1P_2$ physics \label{effectiveChannel}}

We illustrate the relation between the effective $PP$ channel description (i.e., \eqref{H_dark_states} in the main text) and the two channels $P_1P_2$ and $P_2P_1$ description for the three-body problem. 
For the sake of simplicity we neglect weaker off-diagonal vdW interactions $V_{PP,{\rs off}}$.

We consider the Hamiltonian in the following basis: $\ket{SSS}$, $\ket{SP_1P_2}$, $\ket{SP_2P_1}$, $\ket{P_1SP_2}$,$\ket{P_1P_2S}$, $\ket{P_2SP_1}$, and $\ket{P_2P_1S}$. The off-diagonal part of the Hamiltonian:
\beq\left(
\begin{array}{ccccccc}
 0 & \frac{e^{2 i \phi _{2,3}} C_d}{r_{2,3}^3} & \frac{e^{2 i \phi _{2,3}} C_d}{r_{2,3}^3} & \frac{e^{2 i \phi _{1,3}} C_d}{r_{1,3}^3} & \frac{e^{2 i \phi _{1,2}} C_d}{r_{1,2}^3} & \frac{e^{2 i \phi _{1,3}} C_d}{r_{1,3}^3} & \frac{e^{2 i \phi _{1,2}} C_d}{r_{1,2}^3} \\
 \frac{e^{-2 i \phi _{2,3}} C_d}{r_{2,3}^3} & 0 & 0 & \frac{C_{d,1}}{r_{1,2}^3} & 0 & 0 & \frac{C_{d,2}}{r_{1,3}^3} \\
 \frac{e^{-2 i \phi _{2,3}} C_d}{r_{2,3}^3} & 0 & 0 & 0 & \frac{C_{d,1}}{r_{1,3}^3} & \frac{C_{d,2}}{r_{1,2}^3} & 0 \\
 \frac{e^{-2 i \phi _{1,3}} C_d}{r_{1,3}^3} & \frac{C_{d,1}}{r_{1,2}^3} & 0 & 0 & \frac{C_{d,2}}{r_{2,3}^3} & 0 & 0 \\
 \frac{e^{-2 i \phi _{1,2}} C_d}{r_{1,2}^3} & 0 & \frac{C_{d,1}}{r_{1,3}^3} & \frac{C_{d,2}}{r_{2,3}^3} & 0 & 0 & 0 \\
 \frac{e^{-2 i \phi _{1,3}} C_d}{r_{1,3}^3} & 0 & \frac{C_{d,2}}{r_{1,2}^3} & 0 & 0 & 0 & \frac{C_{d,1}}{r_{2,3}^3} \\
 \frac{e^{-2 i \phi _{1,2}} C_d}{r_{1,2}^3} & \frac{C_{d,2}}{r_{1,3}^3} & 0 & 0 & 0 & \frac{C_{d,1}}{r_{2,3}^3} & 0 \\
\end{array}
\right),
\eeq
whereas the diagonal one
\beq
\left(
\begin{array}{c}
 \frac{C_{\text{SS}}}{r_{1,2}^6}+\frac{C_{\text{SS}}}{r_{1,3}^6}+\frac{C_{\text{SS}}}{r_{2,3}^6} \\
 \frac{C_{P_1 P_2}}{r_{2,3}^6}+\Delta _d+\frac{C_{\text{SP}_1}}{r_{1,2}^6}+\frac{C_{\text{SP}_2}}{r_{1,3}^6} \\
 \frac{C_{P_1 P_2}}{r_{2,3}^6}+\Delta _d+\frac{C_{\text{SP}_2}}{r_{1,2}^6}+\frac{C_{\text{SP}_1}}{r_{1,3}^6} \\
 \frac{C_{P_1 P_2}}{r_{1,3}^6}+\Delta _d+\frac{C_{\text{SP}_1}}{r_{1,2}^6}+\frac{C_{\text{SP}_2}}{r_{2,3}^6} \\
 \frac{C_{P_1 P_2}}{r_{1,2}^6}+\Delta _d+\frac{C_{\text{SP}_1}}{r_{1,3}^6}+\frac{C_{\text{SP}_2}}{r_{2,3}^6} \\
 \frac{C_{P_1 P_2}}{r_{1,3}^6}+\Delta _d+\frac{C_{\text{SP}_2}}{r_{1,2}^6}+\frac{C_{\text{SP}_1}}{r_{2,3}^6} \\
 \frac{C_{P_1 P_2}}{r_{1,2}^6}+\Delta _d+\frac{C_{\text{SP}_2}}{r_{1,3}^6}+\frac{C_{\text{SP}_1}}{r_{2,3}^6} \\
\end{array}
\right),
\eeq
where $C_{d}$ denotes dipolar interactions between $SS$ and $P_1P_2$, $C_{d,1}$ between  $SP_1$ and $P_1S$, and $C_{d,2}$ between  $SP_2$ and $P_2S$. Terms without index $_d$ denote vdW interactions.

In order to present the following argument, it is enough to consider only the Hamiltonian elements between five states (out of seven) which we do for the clarity of presentation: 
We rotate the interaction Hamiltonian into the symmetric and asymmetric basis  $\ket{SSS}, \frac{1}{\sqrt{2}}(\ket{SP_1P_2}\pm\ket{SP_2P_1})$, and $\frac{1}{\sqrt{2}}(\ket{P_1SP_2}\pm \ket{P_2SP_1})$.
The off-diagonal terms are:
\beq\left(
\begin{array}{ccccc}
 0 & \frac{\sqrt{2} e^{2 i \phi _{2,3}} C_d}{r_{2,3}^3} & 0 & \frac{\sqrt{2} e^{2 i \phi _{1,3}} C_d}{r_{1,3}^3} & 0 \\
 \frac{\sqrt{2} e^{-2 i \phi _{2,3}} C_d}{r_{2,3}^3} & 0 & \frac{\left(C_{\text{SP}_2}-C_{\text{SP}_1}\right) \left(r_{1,2}^6-r_{1,3}^6\right)}{2 r_{1,2}^6 r_{1,3}^6} & \frac{C_{d,1}+C_{d,2}}{2 r_{1,2}^3} & \frac{C_{d,1}-C_{d,2}}{2 r_{1,2}^3} \\
 0 & \frac{\left(C_{\text{SP}_2}-C_{\text{SP}_1}\right) \left(r_{1,2}^6-r_{1,3}^6\right)}{2 r_{1,2}^6 r_{1,3}^6} & 0 & \frac{C_{d,1}-C_{d,2}}{2 r_{1,2}^3} & \frac{C_{d,1}+C_{d,2}}{2 r_{1,2}^3} \\
 \frac{\sqrt{2} e^{-2 i \phi _{1,3}} C_d}{r_{1,3}^3} & \frac{C_{d,1}+C_{d,2}}{2 r_{1,2}^3} & \frac{C_{d,1}-C_{d,2}}{2 r_{1,2}^3} & 0 & \frac{\left(C_{\text{SP}_2}-C_{\text{SP}_1}\right) \left(r_{1,2}^6-r_{2,3}^6\right)}{2 r_{1,2}^6 r_{2,3}^6} \\
 0 & \frac{C_{d,1}-C_{d,2}}{2 r_{1,2}^3} & \frac{C_{d,1}+C_{d,2}}{2 r_{1,2}^3} & \frac{\left(C_{\text{SP}_2}-C_{\text{SP}_1}\right) \left(r_{1,2}^6-r_{2,3}^6\right)}{2 r_{1,2}^6 r_{2,3}^6} & 0 \\
\end{array}
\right),
\eeq
whereas the diagonal one:
\beq
\left(
\begin{array}{c}
 C_{\text{SS}} \left(\frac{1}{r_{1,3}^6}+\frac{1}{r_{2,3}^6}+\frac{1}{r_{1,2}^6}\right) \\
 \frac{C_{P_1 P_2}}{r_{2,3}^6}+\Delta _d+\frac{1}{2} C_{\text{SP}_1} \left(\frac{1}{r_{1,3}^6}+\frac{1}{r_{1,2}^6}\right)+\frac{1}{2} C_{\text{SP}_2} \left(\frac{1}{r_{1,3}^6}+\frac{1}{r_{1,2}^6}\right) \\
 \frac{C_{P_1 P_2}}{r_{2,3}^6}+\Delta _d+\frac{1}{2} C_{\text{SP}_1} \left(\frac{1}{r_{1,3}^6}+\frac{1}{r_{1,2}^6}\right)+\frac{1}{2} C_{\text{SP}_2} \left(\frac{1}{r_{1,3}^6}+\frac{1}{r_{1,2}^6}\right) \\
 \frac{C_{P_1 P_2}}{r_{1,3}^6}+\Delta _d+\frac{1}{2} C_{\text{SP}_1} \left(\frac{1}{r_{2,3}^6}+\frac{1}{r_{1,2}^6}\right)+\frac{1}{2} C_{\text{SP}_2} \left(\frac{1}{r_{2,3}^6}+\frac{1}{r_{1,2}^6}\right) \\
 \frac{C_{P_1 P_2}}{r_{1,3}^6}+\Delta _d+\frac{1}{2} C_{\text{SP}_1} \left(\frac{1}{r_{2,3}^6}+\frac{1}{r_{1,2}^6}\right)+\frac{1}{2} C_{\text{SP}_2} \left(\frac{1}{r_{2,3}^6}+\frac{1}{r_{1,2}^6}\right) \\
\end{array}
\right).
\eeq
From the off-diagonal terms we see the $\sqrt{2}$ enhancement of the coupling from $SS$ to the symmetric-superposition channel denoted by the $PP$ in the main text.
Coefficients $C_{PP},C_{SP}$ in the main text correspond to the averages of corresponding two-channel quantities.
  
We see that for the generic geometry with $r_{ij}\neq r_{jk}$, decoupling from asymmetric channels requires $C_{SP_1}\approx C_{SP_2}$ and $C_{d,1}\approx C_{d,2}$. %
In our proposal we use $n_1=n$ and $n_2=n-1$ with $n=120\gg 1$ for which $C_{SP_1}/C_{SP_2}\approx 0.98$
$C_{d,1}/C_{d_2}\approx 0.92$. 
This enables us to use the effective single-channel picture to give an intuition behind the multi-body forces. Note that all the numerical results presented in the main text are performed without the single-channel approximation.
Finally, the single-channel picture is valid only for two- and three-body problem.

\section{ Four-body problem\label{fourBody}} 
\begin{figure}[htbp]
\begin{center}
\includegraphics[width=.69 \columnwidth]{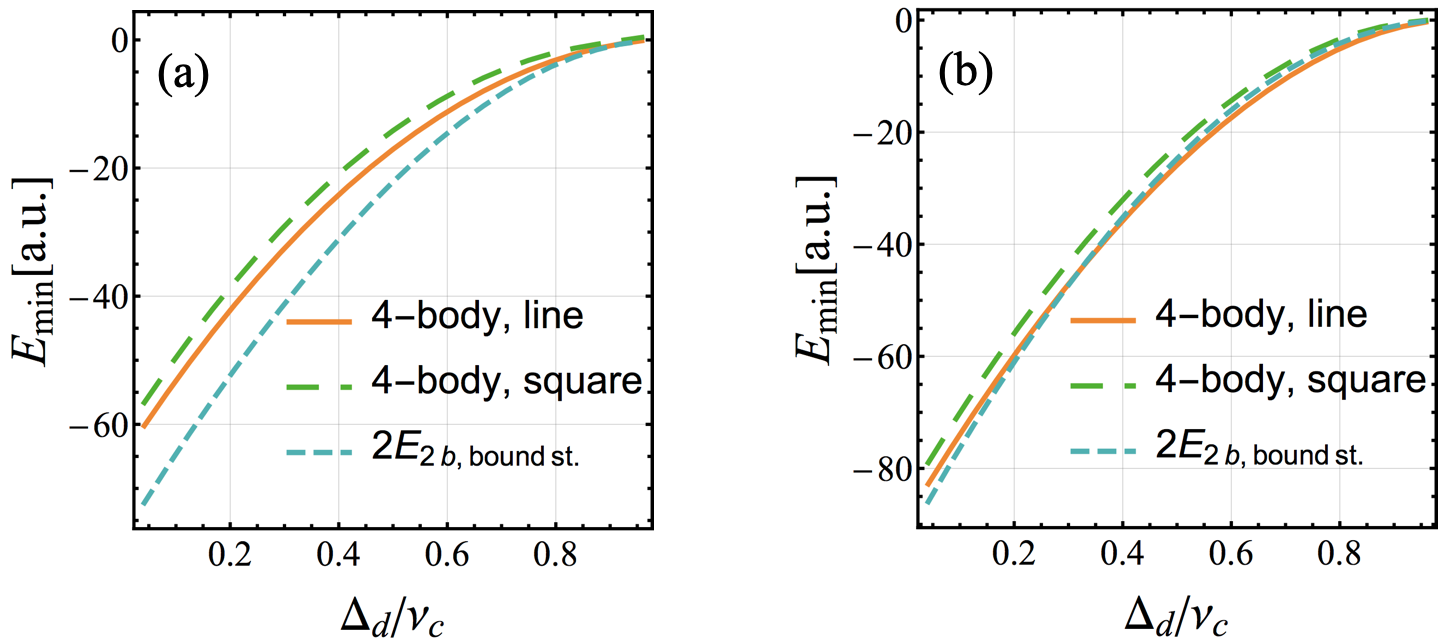}
\caption{Self-consistent solution of Eq.~(6) in the main text describing polaritons in the large-mass limit.
(a-b) Lowest energy  as a function of $\Delta_d$ for line, regular-polygon, and dimer configurations for four bodies. 
Results are shown in (a) for $^{87}$Rb and in (b) for $^{133}$Cs; all of them are for $n_{1},n_2,n$ as in Fig.~2 in the main text. 
}
\label{fig:selfconsistentSupp}
\end{center}
\end{figure}
The four-body problem features additional exotic phenomena due to the strong four-body interactions.
For $^{87}$Rb, the ground state is a configuration consisting of two far-separated dimers~\cite{foot8}, 
see \figref{fig:selfconsistentSupp}(a). %
However for $^{133}$Cs, which has $C_{SP}/C_{PP}$\,$ \approx $\,$ 0.6$ (compared with $C_{SP}/C_{PP}$\,$ \approx $\,$ 1.4$ for $^{87}$Rb) and therefore weaker multi-body forces, 
the ground state configuration depends on $\Delta_d$, \figref{fig:selfconsistentSupp}(b): %
the ground state is a linear configuration
for $\Delta_d/\nu_c\geq 0.3$ and two far-separated dimers for $\Delta_d/\nu_c\leq 0.3$ \cite{foot9}.


%% file: few_body_bound_states.bbl
\begin{thebibliography}{10}
\bibitem{foot2}
One could imagine studying a two-component dark-state polariton (DSP) system,
  where for each Rydberg $S$-state the DSP exists. This, however, is
  experimentally more challenging (requiring more lasers), as well as
  theoretically more complicated, and lacking the simplicity of the
  single-component approach we discuss here.
\bibitem{foot3}
For other choice of $m_J$ for $P$-state, the selection rules would have to be
  modified because for considered magnetic fields we are entering the
  Paschen-Back regime.
\bibitem{foot4}
They are computed by excluding the channels which are taken into account
  explicitly in \eqref{HintForster}.
\bibitem{foot5}
We numerically confirmed that neither the triangular nor the dimer
  configuration is a metastable state.
\bibitem{foot6}
Note that $C_{PP}+C_{PP,{\rs off}}$ in \eqref{eq:Vf} corresponds to $C_{PP}$ in
  the equation above.
\bibitem{foot7}
$W(r)$ is $<0$ because transition dipole elements for the considered levels
  satisfy $\bra{SP_{j}}\hat{d}^-\hat{d}^+\ket{P_jS}<0$.
\bibitem{foot8}
Which we checked is independent of $n$, i.e., $n$ mostly rescales the length
  and energy scales.
\bibitem{foot9}
Note that for the \lineconfiguration, the optimal distance between the edge and
  middle particles is smaller than the distance between the two particles in
  the middle, however, these distances differ by less than $1$\%, and in the
  plots we simply use the average of these distances.
\end{thebibliography}
